\documentclass{optica-article}

\journal{opticajournal} % for journals or Optica Open

\articletype{Research Article}

\definecolor{Blue}{RGB}{15, 98, 254}

\definecolor{Green}{RGB}{25, 134, 32}

\definecolor{teal}{RGB}{26,157,150}

\def\cF{\mathcal{F}}
\def\cV{\mathcal{V}}
\def\cU{\mathcal{U}}
\def\cR{\mathcal{R}}
\def\cL{\mathcal{L}}
\def\cK{\mathcal{K}}
\def\cE{\mathcal{E}}
\def\cP{\mathcal{P}}
\def\cG{\mathcal{G}}

\usepackage{verbatim,bbm,enumitem, amsmath}
\usepackage[absolute]{textpos}

\usepackage[capitalize]{cleveref}
\crefname{section}{Sec.}{Secs.}
\Crefname{section}{Section}{Sections}

\crefrangelabelformat{section}{#3#1#4--#5\crefstripprefix{#1}{#2}#6}
\crefrangelabelformat{figure}{#3#1#4--#5\crefstripprefix{#1}{#2}#6}
\crefrangelabelformat{equation}{(#3#1#4--#5\crefstripprefix{#1}{#2}#6)}

\crefmultiformat{equation}%
{\edef\crefstripprefixinfo{#1}Eqs.~(#2#1#3}%
{,#2\crefstripprefix{\crefstripprefixinfo}{#1}#3)}%
{,#2\crefstripprefix{\crefstripprefixinfo}{#1}#3}%
{,#2\crefstripprefix{\crefstripprefixinfo}{#1}#3)}
\begin{document}

\title{Efficient routing and spectrum allocation in arbitrary flex-grid entanglement networks}

\author{Zachary Goisman,\authormark{1,2,*} Matthew~L. Stevens,\authormark{1,2} Maxwell Goisman, \authormark{2} Taman Truong,\authormark{2,3} Gayane Vardoyan,\authormark{4} Don Towsley,\authormark{4} Nicholas A. Peters,\authormark{5}  Nageswara~S.~V. Rao,\authormark{5} Guoliang Xue,\authormark{2} and Joseph~M. Lukens\authormark{1,2,5}}%,\textdagger}}

\address{\authormark{1}Elmore Family School of Electrical and Computer Engineering and Purdue Quantum Science and Engineering Institute, Purdue University, West Lafayette, Indiana 47907, USA\\
\authormark{2}Ira A. Fulton Schools of Engineering and Research Technology Office, Arizona State University, Tempe, Arizona 85287, USA\\
\authormark{3}Ming Hsieh Department of Electrical and Computer Engineering, University of Southern California, Los Angeles, California 90089, USA\\
\authormark{4}Manning College of Information and Computer Sciences, University of Massachusetts Amherst, Amherst, Massachusetts 01003, USA\\
\authormark{5}Computational Sciences and Engineering Division, Oak Ridge National Laboratory, Oak Ridge, Tennessee 37831, USA}

\email{\authormark{*}zgoisman@purdue.edu}
%\email{\authormark{\textdagger}jlukens@purdue.edu}

\begin{abstract*} 
As practical quantum networks approach large-scale deployment, the need for efficient user-to-user frequency allocation is increasing, yet current approaches only provide partial solutions to the routing and spectrum allocation problem for an arbitrary quantum network.
We address this challenge for repeater-less flex-grid quantum networks based on hyperentangled photons using an efficient three-stage pipeline combining leading tools in classical networking with recent advances in numerical optimization. First, double instantiations of Yen's algorithm obtain low-loss route candidates between each pair of users and the entanglement sources. Second, the advanced process optimizer (APOPT) obtains frequency channel allocations that maximize distribution rates under fidelity constraints. Finally, the constraint programming solver using satisfiability methods (CP-SAT) assigns specific frequency bins to each link, ensuring that there is no contention between frequencies from different sources.
We numerically demonstrate this approach on a representative ring network and a Manhattan incumbent local exchange carrier topology, realizing significant improvements over prior genetic algorithm approaches in speed, accuracy, and scalability. Overall, this pipeline provides an efficient heuristic workflow for optimizing broadband entanglement distribution, applicable to arbitrarily connected quantum networks integrated within the existing lightwave infrastructure.
\end{abstract*}

\begin{textblock}{9.8}(3.1,14.5)
\noindent\fontsize{7}{7}\selectfont \textcolor{black!30}{This manuscript has been co-authored by UT-Battelle, LLC, under contract DE-AC05-00OR22725 with the US Department of Energy (DOE). The US government retains and the publisher, by accepting the article for publication, acknowledges that the US government retains a nonexclusive, paid-up, irrevocable, worldwide license to publish or reproduce the published form of this manuscript, or allow others to do so, for US government purposes. DOE will provide public access to these results of federally sponsored research in accordance with the DOE Public Access Plan (http://energy.gov/downloads/doe-public-access-plan).}
\end{textblock}

\section{Introduction}
\label{sec:introduction}
The creation of secure and reliable quantum network infrastructure is critical to the development and maturity of quantum information science. These networks serve as the backbone for technologies including quantum key distribution (QKD) \cite{ekert_quantum_1991, bennett_quantum_2014, tittel_quantum_2000, scarani_security_2009, pirandola_advances_2020}, distributed quantum sensing~\cite{guo_distributed_2020,zhang_distributed_2021}, and distributed quantum computing \cite{nickerson_topological_2013, monroe_large-scale_2014}. However, in order for these networks to scale to the level of a true quantum internet \cite{kimble_quantum_2008, wehner_quantum_2018}, they must support efficient and contention-free resource allocation.
To this end, quantum networks based on the expansive infrastructure of telecom-band classical lightwave communications are particularly attractive. These classical networks have incorporated flexible-grid wavelength-division multiplexing for dynamic bandwidth allocation to meet user demands~\cite{jinno_spectrum-efficient_2009, christodoulopoulos_elastic_2011, jinno_elastic_2017, chatterjee_routing_2015}---significantly improving spectrum usage by replacing fixed-grid channels with adjustable fine-resolution frequency slots~\cite{noauthor_spectral_2020} and permitting heterogeneous line rates and modulation formats~\cite{christodoulopoulos_elastic_2011, lopez_vizcaino_energy_2012}. Recently, experiments have developed and used flex-grid techniques to dynamically route entangled photons in fiber networks \cite{lingaraju_adaptive_2021, alshowkan_reconfigurable_2021, alshowkan_broadband_2022, alshowkan_resilient_2025}. In their current generation, the primary concern is establishing pairwise entanglement, i.e., maximizing the number of useful photon pairs shared by two nodes.

In classical flex-grid networks, the routing and spectrum allocation (RSA) problem---the standard acronym, not to be confused with Rivest-Shamir-Adleman public-key encryption in quantum cryptographic contexts~\cite{rivest_method_1978}---has been thoroughly studied. Generally, the solution involves integer linear programs~\cite{christodoulopoulos_elastic_2011, chatterjee_routing_2015}, heuristic or metaheuristic algorithms~\cite{chatterjee_routing_2015, walkowiak_routing_2014}, or deep learning~\cite{zhang_online_2022}. Typically, RSA solutions solve both routing and frequency assignment problems simultaneously; however, some methods solve each separately \cite{chatterjee_routing_2015}. Algorithms based on $k$-shortest-path routing~\cite{chatterjee_routing_2015}, fragmentation-aware spectrum assignment~\cite{chatterjee_routing_2015}, dedicated path protection~\cite{walkowiak_routing_2014}, adaptive routing~\cite{alyatama_adaptive_2017},  and evolutionary metaheuristics~\cite{klinkowski_evolutionary_2013, gong_two-population_2012} have all been explored to trade off optimality and computational complexity. Despite classical RSA's position as an obvious stepping stone, quantum RSA must additionally factor in constraints such as entanglement fidelity and source coordination. Several quantum RSA studies have focused on repeater-based architectures aimed at long-distance entanglement distribution and routing~\cite{pant_routing_2019, chakraborty_distributed_2019, shi_concurrent_2024, li_fidelity-guaranteed_2022, halder_optimal_2024, yang_asynchronous_2024}. Others have addressed network protocol layer organization and design~\cite{dahlberg_link_2019, cacciapuoti_multipartite_2024}, or analyzed entanglement-distribution switches and their performance limits in idealized topologies~\cite{vardoyan_stochastic_2021, caleffi_optimal_2017}. However, most of these models treat wavelength as a medium for the entangled bits, rather than as an allocatable resource.

In contrast, today's quantum network testbeds fall in a middle ground that has received comparatively little attention: they lack the deployed quantum repeaters required for the more sophisticated models mentioned above, yet they are also reaching sizes where both resource and lightpath management is becoming an acute need, in terms of both user number and topological flexibility~\cite{Ndousse2019, KleeseVanDam2020, chen_implementation_2021,  chung_design_2022, clark_entanglement_2023,  rakonjac_transmission_2023, bersin2024development, liu_multimode_2024, craddock2024automated, kucera2024demonstration, mckenzie_clock_2024, Stolk2024,  Chapman2024, Baczewski2024, alshowkan_resilient_2025, lukens_hybrid_2025, sena_robust_2025}.
In this more targeted regime of broadband entanglement distribution, the ultimate goal is effectively distributing entangled frequency slots to all users of interest. However, as the number of users grows, determining the optimal allocation faces increasingly complicated instantiations of what has been called the entangled flux allocation (EFA) problem, previously addressed for star topologies with powerful---but slow---genetic algorithms~\cite{alnas_optimal_2022}. While recent research has begun to explore more general multihop quantum versions of the single-source quantum RSA problem~\cite{bali_routing_2025}, no examples to date have combined efficient RSA with multiple entanglement sources and fidelity constraints on arbitrary network topologies. 

Here we introduce and apply a heuristic pipeline combining leading network and optimization tools to solve this problem: (i)~Yen's algorithm for determining low-loss route combinations~\cite{yen_finding_1971}, (ii) the advanced process optimizer (APOPT) for frequency channel allocation~\cite{hedengren_nonlinear_2014, beal_gekko_2018, gunnell_machine_2022}, and (iii) the constraint programming solver using satisfiability methods (CP-SAT) to assign specific frequency bins to lightpaths preventing bandwidth contention~\cite{perron_cp-sat_2025}. After highlighting the basic principles on simple guide configurations, we simulate the pipeline on a seven-source, 24-user ring topology and a three-source, 14-user network modeled after a Manhattan incumbent local exchange carrier (ILEC) \cite{yu_2018, li_2018,bali_routing_2025}. Obtaining the first contentionless solution for both examples in under 0.5~s on a standard laptop computer, our results indicate broad opportunities and prospective scalability for arbitrary repeater-less flex-grid quantum networks.

\section{Theory}
\label{sec:theory}
\subsection{Problem statement}
\label{sec:problem}

In our model, we define each source as a node that generates and sends entangled photon pairs to user nodes, while each user can either drop or forward any frequency slot supported by the network.
An example network can be represented as an arbitrary graph $\cG$ = ($\cV$,$\cE$), where $\cV$ consists of switches and $\cE$ consists of undirected edges. A subset of these switches holds $U$ users $\{u_1,...,u_U\} = \mathcal{V}_{U}$ and the other holds $S$ sources $\{s_1,...,s_S\} = \cV_{S}$, each with $K$ pairs of frequency bins $\{k_{\pm 1},...,k_{\pm K}\}=\cK$. Physically, $\cK$ is determined by the usable entangled photon bandwidth and the frequency bin width. A smaller $K$ therefore corresponds to fewer available spectral resources. Due to energy conservation and the assumed monochromatic pump in the pair generation process, the allocation of bin $k_n$ to one user in a link necessitates the allocation of $k_{-n}$ to the second user.  For concreteness, we reserve ``frequency bin'' for a fundamental frequency channel of a fixed width, whereas ``frequency slot'' denotes an assignment of one or more bins (thus varying in width depending on allocation). The nodes are connected by undirected edges that can sustain photonic propagation, where the users are partitioned into $L\leq\lfloor U/2\rfloor$ pairs $\{(u_{A_1},u_{B_1}),...,(u_{A_L},u_{B_L})\}$ of entangled ``links''---i.e., connections wishing to share Bell pairs. The concatenation of edges through which this link travels is its lightpath. Each edge must not contain identical frequency slots from different sources, as this leads to contention and prevents unambiguous demultiplexing.

\begin{figure}[tb!]
\includegraphics[width=1\linewidth]{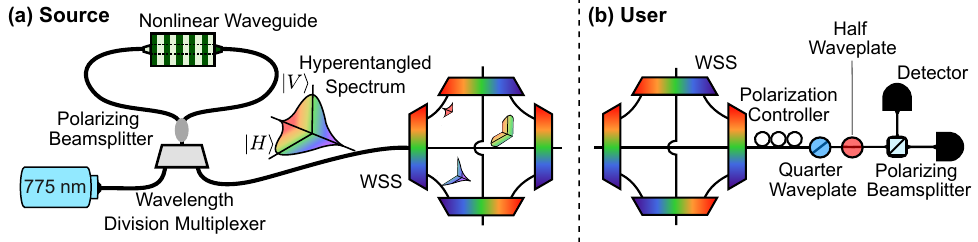}
\caption{Envisioned hardware components for generic (a) source and (b) user nodes, with wavelength-selective switches (WSSs) for arbitrary forwarding of input frequencies to and from adjacent nodes. Here both are shown with $N = 3$ edges, for a total of four WSSs including add-drop. All future topology figures assume these designs, but are simplified to single symbols for clarity.
\label{fig:src_usr}}
\end{figure}

Additionally, we assume that all sources and users can function as switches between adjacent nodes, which for a node with $N$ incident edges can be realized by $N+1$ wavelength-selective switches (WSSs) with at least $N$ output ports each; when appropriately connected, these permit arbitrary forwarding of incoming spectral slices as well as one add-drop connection. \Cref{fig:src_usr} depicts the main components in source and user nodes. Although portrayed for polarization encoding, the model and algorithms we present apply to any fiber-compatible qubit encoding, including time bins~\cite{Xavier2025} and frequency bins~\cite{lu_frequency-bin_2023} with total bandwidth less than the allocated slot.

To focus purely on the RSA-relevant features of our network, we follow \cite{alnas_optimal_2022} in various simplifications, namely: (i) users are entangled only in pairs, (ii) identical quantum states are produced in all frequency bins, and (iii) channel distortion effects are fully compensated. Since each user possesses an add-drop WSS, we assume all other multiplexed channels not allocated to link $\ell$ can be adequately filtered, which allows us to describe each link $\ell\in\{1,...,L\}\equiv[L]$ in isolation of the others. Link $\ell$ is established by a slot comprising $K_\ell$ frequency bins from source $s_{m_\ell}\in\cV_S$ specified by index $m_\ell\in[S]$, where each bin is generated with flux $\mu_{m_\ell}$. The transmissivity $\eta_{A_\ell}$ ($\eta_{B_\ell}$) includes all optical losses from the source to user $u_{A_\ell}$ ($u_{B_\ell}$)  (lightpath through detector), while the noise rate $d_{A_\ell}$ ($d_{B_\ell}$) includes all detection events besides the allocated photons; $\tau$ is the coincidence window, the time interval within which two detection events are classified as a coincidence.

For this model, we can write the fidelity for link $\ell$ as~\cite{alnas_optimal_2022}
\begin{equation}
\label{equ:fidelity}
 \cF_\ell = \frac{1}{4}\left\{ 1 + \frac{3\mu_{m_\ell} K_\ell}{4\tau \left[\frac{\mu_{m_\ell}^2 K_\ell^2}{4} + \left(\frac{d_{A_\ell}}{2\eta_{A_\ell}} + \frac{d_{B_\ell}}{2\eta_{B_\ell}} + \frac{1}{4\tau}\right)\mu_{m_\ell}K_\ell +\frac{d_{A_\ell} d_{B_\ell}}{\eta_{A_\ell}\eta_{B_\ell}}\right]} \right\},
\end{equation}
and the total coincidence rate as
\begin{equation}
\label{equ:rate}
\cR_\ell = 4\tau \left[\frac{\eta_{A_\ell}\eta_{B_\ell}}{4}\mu_{m_\ell}^2 K_\ell^2 + \left(\frac{\eta_{A_\ell}}{2}d_{B_\ell} + \frac{\eta_{B_\ell}}{2} d_{A_\ell} + \frac{\eta_{A_\ell}\eta_{B_\ell}}{4\tau}\right)\mu_{m_\ell}K_\ell + d_{A_\ell} d_{B_\ell}\right].
\end{equation}
More detailed models for entangled-photon states can be found in \cite{truong_developing_2025}, in which it was shown that the preceding equations hold for a variety of user configurations (two or four detectors, threshold or photon-number resolving) in the limits $\mu_{m_\ell}\tau\ll1$, $\eta_{A_\ell,B_\ell}\ll 1$, and  $d_{A_\ell,B_\ell}\tau\ll 1$---all very typical in deployed networks. Note that in contrast to the earlier quantum RSA paper~\cite{alnas_optimal_2022}, the presence of multiple photon sources requires the additional specification $m_\ell$ of the source selected for link $\ell$. In lieu of the full entangled bit error rate (EBR) considered there---defined as \cref{equ:rate} multiplied by the logarithmic negativity---we now focus on the coincidence rate for optimization. Although EBR does quantify both the quality and quantity of entanglement resources, in practical cases we imagine users specifying fidelity requirements up front; hence maximizing $\cR_\ell$ subject to a constraint $\cF_\ell\geq f_\ell$ appears more direct and useful.

Before proceeding further, it is useful to pause and outline the general objectives of the quantum RSA problem in light of \cref{equ:fidelity,equ:rate}. From the network controller's perspective, the overall topology, link pairs $(u_{A_\ell},u_{B_\ell})$, dark count rates $d_{A_\ell}$ and $d_{B_\ell}$, and coincidence window $\tau$ are given as fixed conditions. The quantum RSA problem therefore seeks to (i) assign each link to a source $s_{m_\ell}$ through a pair of routes with efficiencies $\eta_{A_\ell}$ and $\eta_{B_\ell}$; (ii) adjust the bin flux $\mu_{m_\ell}$ as desired and allocate $K_\ell\leq K$ of the source's pairs of frequency bins to the link; and (iii) repeat over all $L$ links avoiding bandwidth contention in all network edge. Any RSA algorithm must furnish $S$ source flux values $\{\mu_1,...,\mu_S\}$; $L$ frequency slot widths  $\{K_1,...,K_L\}$ and center frequencies; and $L$ pairs of lightpaths with efficiency products $\{\eta_{A_1}\eta_{B_1}, ..., \eta_{A_L}\eta_{B_L}\}$. 

Therefore, we summarize our complete global RSA problem formally as the following:
\begin{enumerate}
\item Begin with a graph $\cG = (\cV, \cE)$ consisting of $S+U$ total vertices split into two types: $S$ sources $\{s_1,...,s_S\}=\cV_S$ and $U$ users $\{u_1,...,u_U\}=\cV_U$, where $\cV_S \cup \cV_U = \cV$. These vertices are connected by $E$ undirected edges $\cE = \{e_1,...,e_E\} \subseteq \{(v,w):v,w \in \cV_S \cup \cV_U,v \neq w \}$, each of which is characterized by an efficiency parameter $h_{e}\in(0,1]$.

\item The users wish to establish a maximum of $L\leq \lfloor U/2\rfloor$ user-disjoint pairwise links described by the set $\{(u_{A_1},u_{B_1}),...,(u_{A_L},u_{B_L})\}=\cL$, where each link $\ell\in[L]$ possesses an ``Alice'' $u_{A_\ell}$ and ``Bob'' $u_{B_\ell}$ and requests entanglement of fidelity exceeding some threshold $\cF_\ell\geq f_\ell$.

\item Each source $s_m$ ($m\in[S]$) possesses $K$ pairs of entangled frequency bins $\cK=\{k_{\pm 1},...,k_{\pm K}\}$, each produced at the same flux $\mu_m$. Under this definition, $|\cK|=2K$. The set of links serviced by source $s_m$ is denoted by $\cL_m\subset \cL$.

\item The quantum RSA problem seeks to determine the following:
\begin{enumerate}
\item Channel flux value $\mu_m$ for each source $m\in[S]$.

\item Source assignment $m_\ell\in[S]$ and lightpaths to both users in link $\ell\in[L]$. Let $\cP_{A_\ell}\subseteq\mathcal E$ be the set of edges comprising a path from $s_{m_\ell}$ to $u_{A_\ell}$, and let $\cP_{B_\ell}\subseteq\mathcal E$ be the set of edges comprising a path from $s_{m_\ell}$ to $u_{B_\ell}$.
These lightpaths have efficiencies
$\eta_{A_\ell}=\prod_{e\in \cP_{A_\ell}} h_e$ and $\eta_{B_\ell}=\prod_{e\in \cP_{B_\ell}} h_e$.

\item Spectral assignment for each link $\ell\in[L]$, where $\cK_{A_\ell}\subset\cK$ bins are assigned to Alice and $\cK_{B_\ell}\subset\cK$ to Bob, which are energy-matched: i.e., for every $k_n\in\cK_{A_\ell}$, it must be the case that $k_{-n}\in\cK_{B_\ell}$; hence $|\cK_{A_\ell}|=|\cK_{B_\ell}|\equiv K_\ell$.
\end{enumerate}

The goal is to maximize the chosen definition of network utility $\cU$ subject to the following constraints:
\begin{enumerate}[label=(C\arabic*)]
\item \textbf{Fidelity:} $\cF_\ell\geq f_\ell \; \forall \; \ell\in[L]$.
\item \textbf{Per-source capacity:} $\displaystyle\sum_{(u_{A_\ell},u_{B_\ell})\in\cL_m} K_\ell \leq K \; \forall \; m\in[S]$.
\item \textbf{Contention-free distribution:} $\cK_{\alpha_\ell} \cap \cK_{\beta_{\ell'}}=\emptyset$ for all $\ell\neq\ell'$ and $\alpha,\beta\in\{A,B\}$ whose paths $\cP_{\alpha_\ell}$ and $\cP_{\beta_{\ell'}}$ share an edge.
\end{enumerate}
\end{enumerate}

\subsection{Example topologies}
\label{sec:topology}
\begin{figure}[tb!]
\includegraphics[width=1\linewidth]{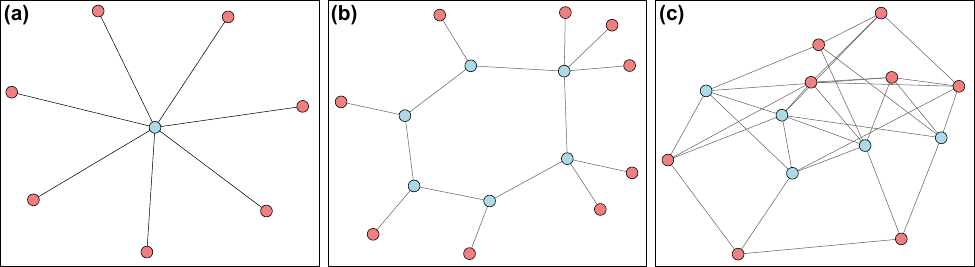}
\caption{Example network topologies solvable by the quantum RSA workflow: (a)~star, (b) ring, and (c) densely connected. Sources are represented by blue circles, users by red circles.
\label{fig:topology}}
\end{figure}

One of the major enhancements of our approach compared to previous related work~\cite{alnas_optimal_2022,bali_routing_2025} is the consideration of both multiple sources and multihop routing. \Cref{fig:topology} shows examples of arbitrary ring, star, and dense topologies, generated by the NetworkX package in Python~\cite{hagberg_exploring_2008}, which all work with our routing optimizer. Star topologies consist of just one source with all users connected to it [\cref{fig:topology}(a)]. \Cref{fig:topology}(b) depicts a ring topology where only sources make up the inner ring. All user nodes are then attached directly to only one source as in the star case. Lastly, \cref{fig:topology}(c) shows a densely connected network topology, controlled by a density factor $D\in(0,1]$, defined as the ratio of the number of edges $|\cE|$ to the maximum possible $|\cV|(|\cV|-1)/2$. In this specific example, $D=0.4$.

\section{Methods}
\label{sec:methods}
\begin{figure}[tb!]
\includegraphics[width=1\linewidth]{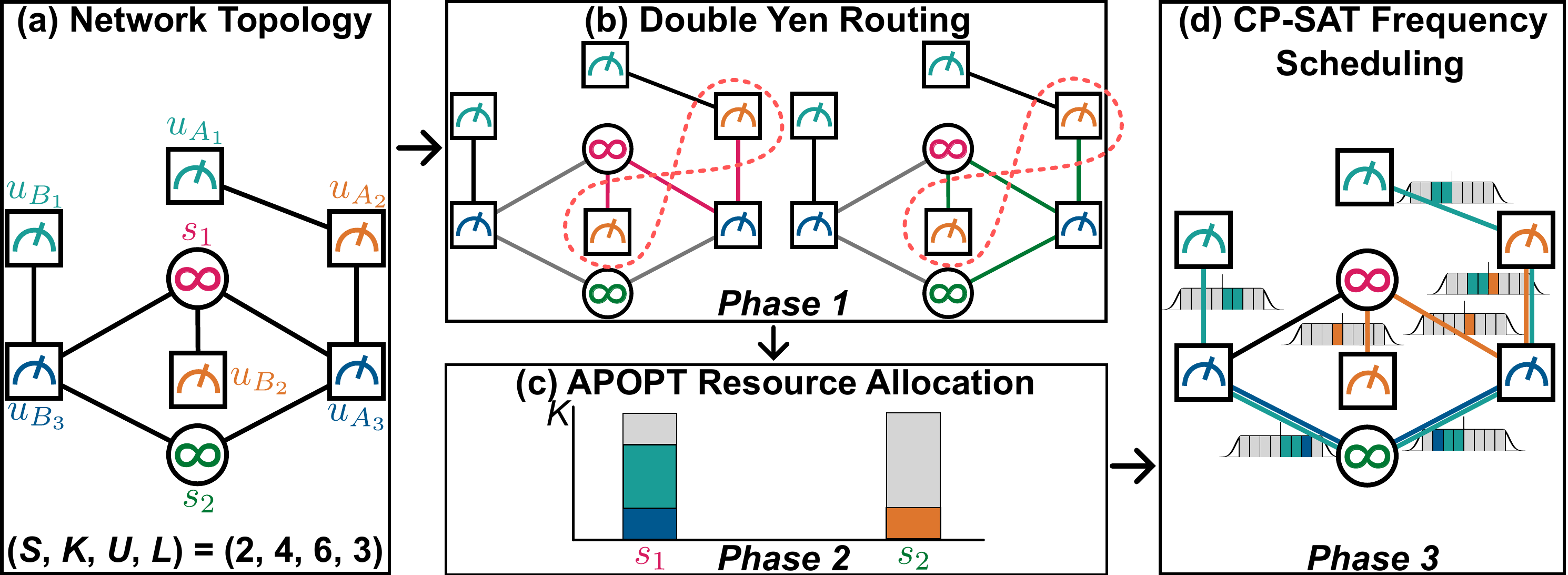}
\caption{Overview of proposed quantum RSA pipeline. (a) Example network topology with two sources (infinity signs within a circle) and six users (measure operation squares). Links are represented by two users with the same color: $\ell=1$ is teal, $\ell=2$ orange, and $\ell=3$ blue. (b) Phase 1: double Yen routing. This image shows one possible pair of paths from each source to the two users comprising the orange link. (c) Phase 2: APOPT spectrum allocation. From the pink $s_1$ source, the teal link is allocated two channels and the blue link one channel, while the orange link receives a single channel from the green $s_2$ source. For both sources, a total of $K=4$ bin pairs are assumed. (d) Phase 3: CP-SAT solver scheduling the allocated frequency slots to avoid contention. The eight-channel spectrum on each edge depicts the used frequency bins (gray is unassigned).
\label{fig:main}}
\end{figure}

Our approach is motivated by the computational complexity of the problem described in \cref{sec:problem}, which is NP-hard because the allocation subproblem is NP-hard. On its own, the EFA problem \cite{alnas_optimal_2022}---the special case of a star topology and one source [e.g., \cref{fig:topology}(a)]---has been shown to possess the form of the multiple subset sum problem (MSSP) \cite{caprara_multiple_2000}, which is strongly NP-hard. Additionally, once lightpaths and the number of frequency bins are assigned, contention-free frequency-bin scheduling is NP-complete~\cite{fernandes_da_silva_framework_2022}. In light of these complexity-based challenges,  we propose a heuristic pipeline that divides quantum RSA into three phases: lowest-loss pathfinding (Phase 1), spectrum allocation (Phase 2), and frequency scheduling (Phase 3). This segmentation allows us to exploit the well-known Yen algorithm for Phase 1~\cite{yen_finding_1971}, which can be solved in polynomial time and thereby reduces the size of NP-hard phases into more manageable chunks. Our heuristic pipeline is depicted for a simple example in \cref{fig:main} and is explained in depth in this section.

The example topology [\cref{fig:main}(a)] comprises a densely connected network with $(S,K,U,L)=(2,4,6,3)$; i.e., two sources with four pairs of frequency bins each and six users desiring three pairwise entangled links. Each circle denotes a source  assumed to contain the components in \cref{fig:src_usr}(a), while each square represents a user with the components in \cref{fig:src_usr}(b).

\subsection{Phase 1: routing}
\label{subsec:routing}
With an arbitrary topology provided, any pipeline must find the best routes and optimal spectrum allocation for each link. Although the full problem is strongly NP-hard, significantly restricted versions are more tractable. For instance, we can readily determine an optimal route for a \emph{predetermined} spectrum allocation, because the rates for each potential route are easily calculable. We can also determine an optimal spectrum allocation for a fixed set of link routes, as we can allocate frequencies based on preventing bandwidth contention. Unfortunately, it is significantly more nuanced if neither the allocation nor routing begins at its respective optimum, since the two depend on each other. To determine the best routes, one needs to evaluate the utility for all links, but to get this utility, one must first provide a spectrum allocation that depends on the route.

In our case, we confront this circularity by addressing routing first, completely independent of spectrum allocation. 
Specifically, in \textit{Phase~1}, we apply Yen's algorithm \cite{yen_finding_1971} to find the $N$ lowest-loss paths (where $N$ is a small user-specified number)  from every source $s_m$ to each user $(u_{A_\ell},u_{B_\ell})$ totaling to $N^2$ combinations for a specific source-link combination, which we dub ``double Yen.''
Taking $N>1$ increases the size of the routing search and is not intrinsically required ($N=1$ is a perfectly valid choice for Phase 1), yet it furnishes backup routes in the event that a contention-free solution cannot be obtained in the subsequent two phases. Empirically, we found $N\leq4$ sufficient for all examples in \cref{sec:results} below, but any $N$ can be chosen as desired in a network scenario. We preliminarily assign each link to the source-lightpath combination $\cP_{A_\ell}$ and $\cP_{B_\ell}$ with the maximum combined efficiency $\eta_{A_\ell}\eta_{B_\ell}$ and rank all $(N^2 S)^L$ lightpath combinations (every source-link assignment and double Yen permutation) according to a global efficiency defined as 
\begin{equation}
\label{eq:totEff}
\eta_\text{tot} = \prod_{\ell\in[L]}\eta_{A_\ell}\eta_{B_\ell}.
\end{equation}

\Cref{fig:main}(b) depicts an example of the $N = 1$ case (chosen for visual simplicity), where the pink and green lines denote the lowest-loss pair of paths for $(u_{A_2},u_{B_2})$ to the sources $s_1$ and $s_2$, respectively. Hence, Phase~1 returns the assigned source $m_\ell$, the corresponding routes $\cP_{A_\ell}$ and $\cP_{B_\ell}$, and the total efficiencies $\eta_{A_\ell}$ and $\eta_{B_\ell}$ for each link $\ell\in[L]$, using the aggregate routing assignments that maximize \cref{eq:totEff} for this first attempt. 

\subsection{Phase 2: spectrum allocation}
\label{subsec:resource_allocation}
With a preliminary assignment of sources and routes from Phase~1, resulting in the association of links to sources and total efficiencies $\eta_{A_\ell}$ and $\eta_{B_\ell}$ of these paths for a chosen route, \textit{Phase~2} solves the simpler EFA problem~\cite{alnas_optimal_2022} for each source individually, determining the source flux $\mu_m$ and number of channels $K_\ell$ each link  should receive.  
Since the goal is rate maximization via \cref{equ:rate}, we seek to increase $\mu_{m_\ell}K_\ell$ as much as possible until hitting the specified constraint $f_\ell$, the maximum of which we can analytically calculate from \cref{equ:fidelity}, thus providing ``infinite-resource'' flux targets for each link: 

\begin{equation}
\label{equ:upper_bound}
\mu_{m_\ell}K_\ell\Big|_\infty = \frac{1}{2\tau}\left(a+\sqrt{a^2-\frac{16\tau^2 d_{A_\ell}d_{B_\ell}}{\eta_{A_\ell}\eta_{B_\ell}}}\right); \quad
a=\frac{1-f_\ell}{f_\ell-\frac{1}{4}}-\frac{2\tau d_{A_\ell}}{\eta_{A_\ell}}-\frac{2\tau d_{B_\ell}}{\eta_{B_\ell}},
\end{equation}
assuming a real solution exists.
We then define $\cR_\ell^\infty$ as the value of \cref{equ:rate} using these values; $\cR_\ell^\infty$ therefore denotes the maximum possible coincidence rate under the fidelity constraint $f_\ell$ for the link's best path from Phase 1.
Accordingly, if the number of usable channels per source $K$ approaches infinity, all links can approach their maximum rate with the channel flux $\mu_{m_\ell}$ fixed at some suitably small value. As argued in \cite{alnas_optimal_2022}, the EFA problem is a variation on the MSSP \cite{caprara_multiple_2000}, a type of multiple knapsack problem that seeks to maximize the total value carried by  identically sized knapsacks, through filling them with variously weighted items. 

\begin{figure}[tb!]
\centering
\includegraphics[width=1\linewidth]{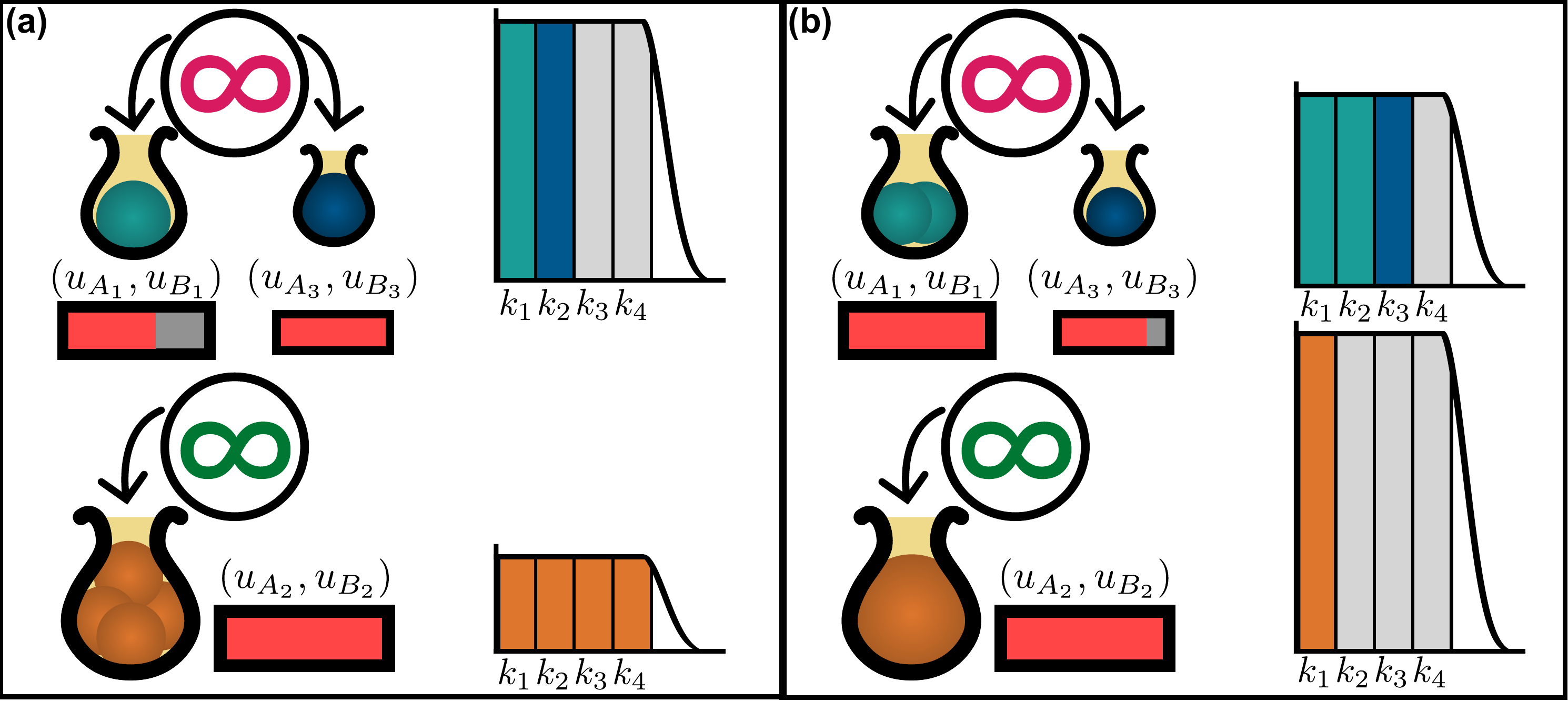}
\caption{Depiction of the EFA problem (Phase 2 of \cref{fig:main}) in terms of a knapsack problem. (a) An example inefficient case, where source $s_1$ allocates a channel that maximally fills the knapsack of the blue link $\ell=3$, preventing it from further filling the teal link $\ell=1$, while source $s_2$ assigns all four channel pairs to its single orange link $\ell=2$. (b) An improved solution in which the flux of $s_1$ is reduced such that three channels spread across its two links, and the flux of $s_2$ is increased to require only one channel instead of all four. In all examples, the size of the knapsack and corresponding slider symbolizes the maximum flux each link can receive, the size of the balls stand in for the flux of an individual channel, and frequency-bin assignments for each source are shown as a ``half-spectrum'' since the lower-frequency half is the mirror image.
\label{fig:knapsack}}
\end{figure}

This knapsack interpretation for the topology in \cref{fig:main} is depicted in \cref{fig:knapsack}, where links $\ell=1$ (teal) and $\ell=3$ (blue) are assigned to $s_1$ and $\ell=2$ (orange) to $s_2$. The size of each knapsack corresponds to the maximum flux a given link can accept, with each ball size proportional to the flux for a specific channel. By lowering the flux of the individual channels for source $s_1$ from panel (a) to (b), more overall flux can be allocated to its two links $\ell=1$ and $\ell=3$. In contrast, by increasing the flux of source $s_2$ from (a) to (b), the link $\ell=2$ knapsack can be fully filled with fewer channels, leaving unassigned frequency bins for future use.

Whereas Phase~1 possesses an unambiguous objective for optimality in terms of combined efficiency $\eta_{A_\ell}\eta_{B_\ell}$ for each $\ell$, the optimal EFA solution depends on the choice of cost function, which can vary depending on application. Using \cref{equ:rate}, $\cF_\ell$ from \cref{equ:fidelity} can be written as 
\begin{equation}
\label{equ:solvefidelity}
      \cF_\ell=\frac{1}{4}\left(1+\frac{3\eta_{A_\ell}\eta_{B_\ell}\mu_{m_\ell}K_\ell}{\cR_\ell}\right).
\end{equation}
As $\cR_\ell$ depends quadratically on $\mu_{m_\ell}K_\ell$ [\cref{equ:rate}], there exists an inherent tradeoff between fidelity and rate; namely, any rate $\cR_\ell>\cR_\ell^\infty$ violates the fidelity constraint $\cF_\ell\geq f_\ell$. Hence, we aim to maximize the rate until the fidelity hits its specified limit.

Inspired by the history of network utility in classical networking~\cite{kelly_rate_1998, low_optimization_1999} and recent adaptations to the quantum domain~\cite{vardoyan_quantum_2023}, we select a utility defined
as the sum of logarithmic rates, namely~\cite{vardoyan_quantum_2023}
\begin{equation}
\label{equ:sumrate}
      \cU_m = \sum_{(u_{A_\ell},u_{B_\ell})\in\cL_m} \log_{10}\cR_\ell
\end{equation}
for source $s_m$, where the sum is performed over all links currently assigned to this source, formally defined as the set $\cL_m=\{(u_{A_\ell},u_{B_\ell})\in\cL : m_\ell=m\}$. The global utility $\cU$ follows as the sum of the per-source utilities:
\begin{equation}
\label{eq:globalU}
\cU=\sum_{m\in[S]}\cU_m=\sum_{\ell\in[L]}\log_{10}\cR_\ell,
\end{equation}
where we can again define an infinite-resource value $\cU_\infty$ for the case $\cR_\ell=\cR_\ell^\infty\;\forall\;\ell\in[L]$.
This logarithmic utility expresses quantitatively the qualitative desire to prevent smaller links from being left unfilled, while still prioritizing high overall rates.
Consequently, we can write the EFA problem for source $s_m$ as 
\begin{equation}
\label{equ:minlp}
\begin{aligned}
(\hat{\mu}_m,\{\hat{K}_\ell\}_{\ell\in\mathcal{L}_m})
&=
\operatorname*{arg\,max}_{\mu_m,\{K_\ell\}_{\ell\in\mathcal{L}_m}}
\cU_m
\\
&\text{s.t.}\quad
\cF_\ell \ge f_\ell,
\quad\sum_{\ell\in\mathcal{L}_m} K_\ell \le K,
\quad\mu_m \in\mathbb{R}_{>0},\quad
K_\ell \in \mathbb{N}_1.
%\quad \forall\, \ell\in\mathcal{L}_m .
\end{aligned}
\end{equation}
We note that the base chosen for the logarithm has no impact on the solution to the EFA optimization problem expressed by \cref{equ:sumrate}, but does influence quantitative interpretations of $\cU_m$. For example, the base-10 logarithm in \cref{equ:sumrate} implies that  
a reduction in utility of $-1$ for a given link compared to the infinite-resource limit corresponds to $\cR_\ell$ equal to 10\% of its maximum.

In light of the constraints in \cref{equ:minlp}, spectrum allocation takes the form of a mixed-integer nonlinear programming (MINLP) problem, for which we have found   
APOPT \cite{hedengren_nonlinear_2014, beal_gekko_2018, gunnell_machine_2022} to solve extremely efficiently. APOPT was specifically developed for large-scale nonlinear and dynamic optimization problems arising in science and engineering.
In our application, APOPT uses a branch-and-bound method to split the MINLP problem into a sequence of continuous nonlinear programming subproblems that can be solved with an active-set sequential quadratic programming method, which considers local quadratic approximations. Upon completion of its $S$ instantiations of the EFA problem (one for each source), Phase 2 will furnish all source fluxes $\mu_m$ and slot widths $K_\ell$.

\subsection{Phase 3: frequency scheduling}
\label{subsec:frequencypathing}

Following Phase~2 of the pipeline, each link will have been assigned a specific source $m_\ell$, two lightpaths $\cP_{A_\ell}$ and $\cP_{B_\ell}$, and number of allocated channel pairs $K_\ell$---fully sufficient to complete RSA in a single-source network ($S=1$). However, the presence of multiple sources requires additionally that the \emph{specific} frequency bins comprising the slots $\cK_{A_\ell}$ and $\cK_{B_\ell}$ be assigned and checked for contention---i.e., identical physical frequencies from different sources attempting to traverse the same network edge. \emph{Phase~3} is therefore concerned with 
determining specific frequency allocations that avoid this contention, or else learn that certain path combinations from Phase~1 are not viable. Since photon pairs have complementary frequencies, we can access more potential scheduling options than in the classical RSA problem: e.g., both routes of a pair  $\cP_{A_\ell}$ and $\cP_{B_\ell}$ can traverse the same edges, and any frequency bins can be swapped between Alice's and Bob's sets ($\cK_{A_\ell}$ and $\cK_{B_\ell}$) with no change to rate or fidelity [under the assumptions behind \cref{equ:fidelity,equ:rate}].

We enlist CP-SAT to handle backtracking in the event of contention  \cite{perron_cp-sat_2025}. This choice is motivated by the discrete nature of bandwidth contention and channel scheduling, where CP-SAT was designed for integer-constrained combinatorial problems, and can efficiently prune infeasible assignments. If the optimizer cannot resolve all conflicts, we select the next most-efficient combination of double Yen routes found in Phase 1, as ranked by the value of $\eta_\text{tot}$ in \cref{eq:totEff}. We then rerun APOPT in Phase 2 to find the bandwidth allocation for these new lightpaths and recheck for contention, repeating  until a contention-free RSA solution is obtained.

\section{Results}
\label{sec:results}
\subsection{Highly asymmetric contention example}
\label{subsec:contention}
\begin{figure}[tb!]
\centering
\includegraphics[width=1\linewidth]{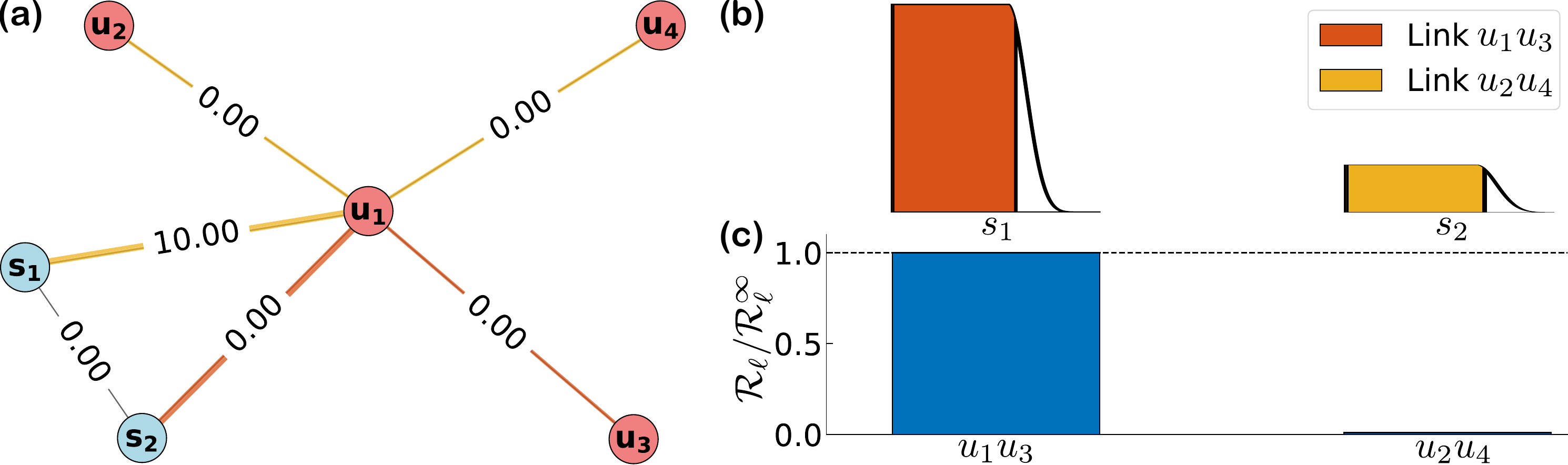}
\caption{Solution for a specifically designed network to induce bandwidth contention with $S=2$ sources, $K=1$ channel, $U=4$ users, and $L=2$ desired entangled links. (a)~Topology and source-to-user lightpaths for each entangled link, with edges labeled by their losses in dB. (b)~Channels allocated to each entangled link by source, where the bar plot height is proportional to the flux each source provides. (c)~Individual link rates $\cR_\ell/\cR_{\ell}^{\infty}$ normalized to infinite-resource maxima.
\label{fig:contention}}
\end{figure}

\Cref{fig:contention} shows a deliberately pathological network with $(S,K,U,L)=(2,1,4,2)$ designed to highlight the challenge of contention and the solution obtained by our method. We set all dark count rates to $0~\text{s}^{-1}$, minimum fidelities to $f_1 = 0.5$ and $f_2 = 0.75$, and $\tau=1$~s. Link $\ell = 1$ corresponds to users $(u_1,u_3)$ and link $\ell = 2$ to $(u_2,u_4)$. Here, all edges are assigned losses of 0~dB except for the 10~dB path from $s_1$ to $u_1$, thus making the path efficiencies $\eta = 1$ or $0.1$. Consequently, both links $(u_1,u_3)$ and $(u_2,u_4)$ attempt to avoid this edge, yet because each source only has one frequency bin pair, this avoidance leads to bandwidth contention on the $(s_2,u_1)$ edge, even after assigning each to a different source per the constraint~(C2). It is not until the 20th route configuration [ordered by $\eta_\text{tot}$ in \cref{eq:totEff}] that the pipeline obtains a contentionless solution satisfying (C3). This completes in 0.241 s, and reaches $\cU = -1.35$ compared to the infinite-resource upper bound $\cU_\infty=0.653$.

As highlighted in \cref{fig:contention}(b) the $(u_1,u_3)$ link attains its maximum possible rate exactly, while $(u_2,u_4)$ is only at 50.5\%. This reduction is expected, as in order to best avoid contention, we must force the second link through the high-loss edge, despite the existence of the lower-loss route. Since each source can independently optimize its flux for its single link, the optimizer can easily reach the minimum fidelity requirement (C1) for each link by choosing channel fluxes of $2~\text{s}^{-1}$ and $0.5~\text{s}^{-1}$ for $s_1 \text{ and } s_2$, respectively. Of course, considering sources with only one available pair of frequency bins is intentionally contrived: the entire concept of the quantum RSA problem is inspired by the availability of quantum light sources with many (even hundreds) of frequency bins~\cite{alshowkan_broadband_2022}. Nevertheless, this edge case highlights the success of the pipeline in addressing even extreme scenarios.

\subsection{Complex network examples}
\begin{figure}[tb!]
\centering
\includegraphics[width=1\linewidth]{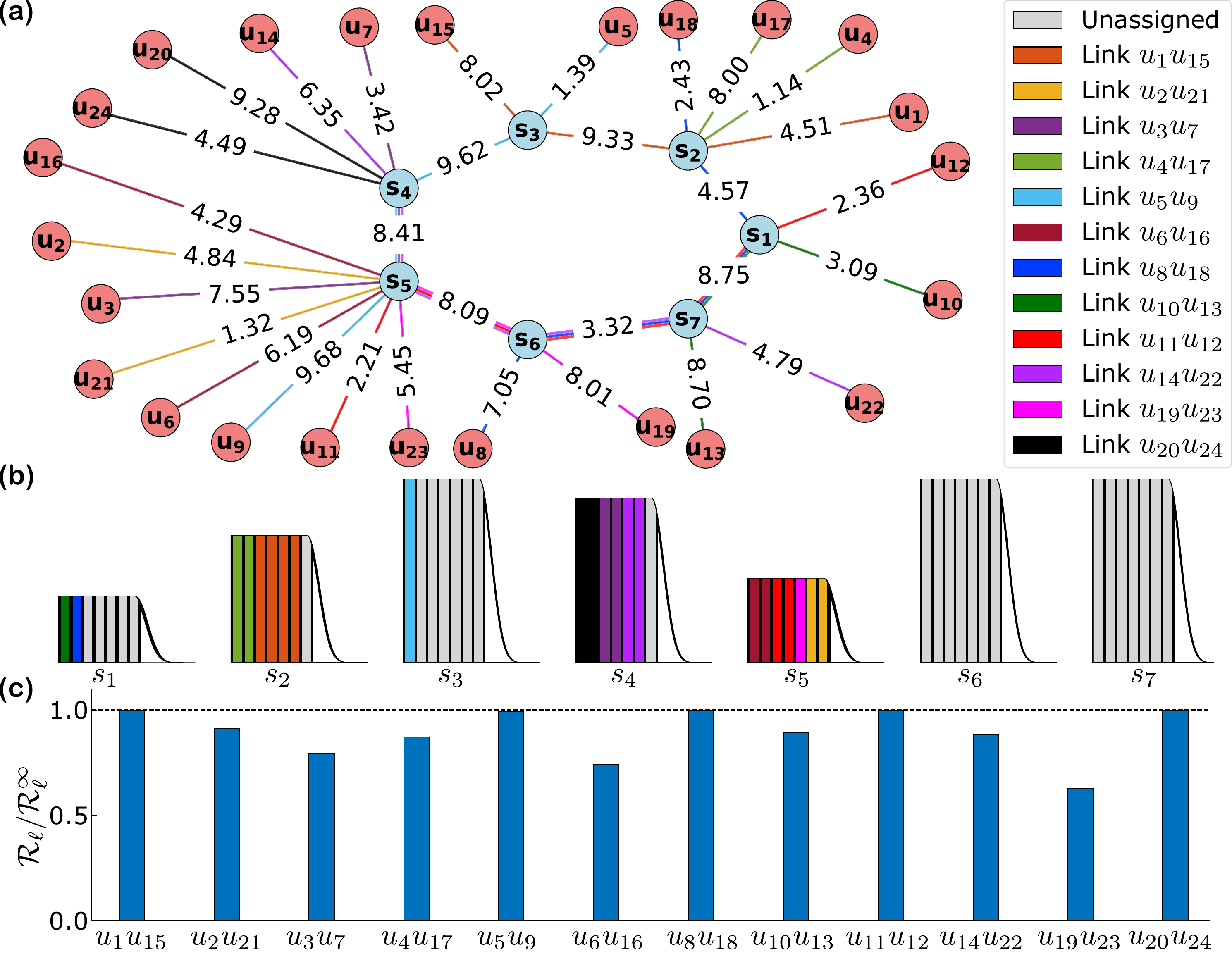}
\caption{Solution for a randomly generated network with $S=7$ sources, $K=7$ channels, $U=24$ users, and $L=12$ desired entangled links. (a)~Topology and source-to-user lightpaths for each entangled link, with edges labeled by their losses in dB. (b)~Channels allocated to each entangled link by source, where the bar plot height is proportional to the flux each source provides. (c)~Individual link rates $\cR_\ell/\cR_{\ell}^{\infty}$ normalized to infinite-resource maxima.
\label{fig:combined_results_ring}}
\end{figure}

To demonstrate its effectiveness at routing and allocating resources to all links, we run the entire pipeline on two example networks. For a comparison to the GA-based approach of \cite{alnas_optimal_2022}, see  \cref{subsec:comparison}. Unless stated otherwise, for each of the tested topologies we draw dark count rates ($d_{A_\ell},d_{B_\ell}$) uniformly at random from the interval $[100~\text{s}^{-1}, 1000~\text{s}^{-1}]$, minimum fidelities ($f_\ell$) from the interval $[0.9,0.95]$, and edge losses ($-10\log_{10} \eta_{A_\ell}, -10\log_{10}\eta_{B_\ell}$) from the interval $[1~\text{dB}, 10~\text{dB}]$; $\tau=1$~ns throughout.
For ring topologies, we have found that using any of the $N-1$ next lowest-loss paths dramatically increases the potential for bandwidth contention, making $N=1$ a reasonable simplification for Phase 1. Taking $N=1$ for the ring example in \cref{fig:combined_results_ring} with $(S,K,U,L)=(7,7,24,12)$, we % but now with $N=1$. Additionally, edge losses are drawn randomly from [1,10]~dB. %Edge losses are also drawn randomly from [1,5]~dB between users and sources, and drawn randomly from [5,10]~dB.
obtain the first contentionless RSA solution in
0.498~s, reaching utility $\cU= 73.2$ compared to the $K\rightarrow\infty$ upper bound of $\mathcal{U}_\infty=73.9$. This solution achieves individual link rates on average equal to 89.2\% of their infinite-resource maxima [\cref{fig:combined_results_ring}(c)].

\begin{figure}[tb!]
\centering
\includegraphics[width=1\linewidth]{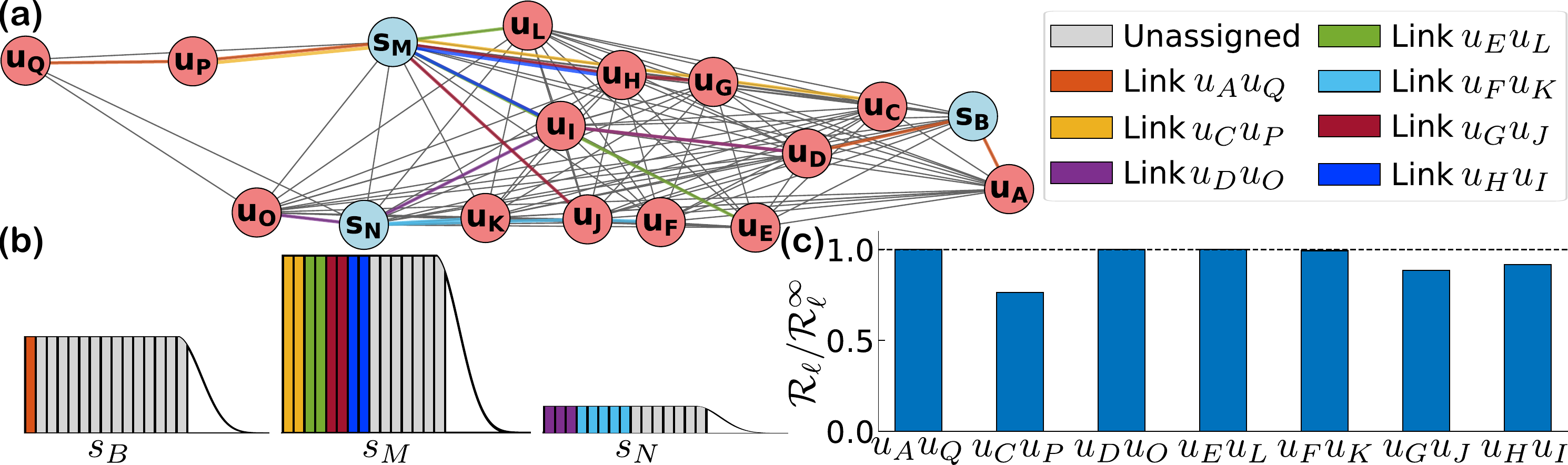}
\caption{Solution for Manhattan ILEC network from \cite{bali_routing_2025, yu_2018, li_2018} with $S=3$ sources, $K=15$ channels, $U=14$ users, and $L=7$ desired entangled links. (a)~Topology and source-to-user lightpaths for each entangled link, with edge weights obtained by converting the reported intersite distances in Table 1 of \cite{bali_routing_2025} in dB using an effective attenuation coefficient of $1~\mathrm{dB/km}$. The node positions are shown as a schematic layout of the island oriented so that north is pointed rightwards, rather than exact physical geographic coordinates. (b)~Channels allocated to each entangled link by source. The bar plot height is proportional to the flux each source provides. (c)~Individual normalized rates $\cR_\ell/\cR_{\ell}^{\infty}$ relative to infinite-resource maxima rates.
\label{fig:combined_results_dense}}
\end{figure}

\Cref{fig:combined_results_dense} shows results on a Manhattan ILEC topology with density $D=0.81$ used by \cite{bali_routing_2025} and originally derived from \cite{yu_2018,li_2018}. Since deployed fiber lengths are unavailable, we follow \cite{bali_routing_2025} using ``as the crow flies'' distances with $1~\text{dB km}^{-1}$ fiber attenuation, $N = 4$ for the pathfinding in Phase~1, and $(S,K,U,L)=(3,15,14,7)$. The pipeline takes 0.283~s to obtain a contentionless RSA solution with utility $\cU= 51.1$, compared to the infinite-resource bound $\mathcal{U}_\infty=51.3$, for an average link rate of 93.7\% compared to its infinite-resource ceiling [\cref{fig:combined_results_dense}(c)].

In order to minimize total runtime, the pipeline as presented so far concludes upon identification of the \emph{first} contentionless solution in Phase 3. However, this first solution is not guaranteed to be the \emph{global} optimum, as it is tied to a specific collection of double Yen routes in Phase 1 that may not correspond to the utility-maximizing configuration over all possible RSA settings. The motivation behind prioritizing routes according to the efficiency defined in \cref{eq:totEff} is precisely to avoid settling on lightpaths significantly different from the global optimum, but this ultimately represents little more than an informed guess.

To explore the effectiveness of the lowest-loss ordering, we perform a semi-exhaustive search over up to 10,000 possible source and routing configurations in Phase 1 for the networks in \cref{fig:combined_results_ring,fig:combined_results_dense}: \cref{tab:optima} summarizes our findings, highlighting the total number of route combinations and runtime for the semi-exhaustive search; the route number and runtime to reach the first contentionless solution, along with its utility; and the route number, runtime, and utility of the best solution. Overall, we find that this first selection of routes is extremely close to the best network utility we could find. Prospective network route combinations balloon in number with the addition of more links and sources, and it is not practical to run the approximately 13.8 billion possible combinations for Fig. \ref{fig:combined_results_ring} or 587 billion for Fig. \ref{fig:combined_results_dense}. Taking the first possible solution also enables us to solve significantly larger networks on the order of seconds rather than hours.

\begin{table}[tb!]
    \centering
    \footnotesize
    \begin{tabular}{|l|cc|ccc|ccc|}
    \hline
            &  \multicolumn{2}{c|}{\textit{Semi-exhaustive Search}} & \multicolumn{3}{c|}{\textit{First solution}} & \multicolumn{3}{c|}{\textit{Best semi-exhaustive solution}}  \\
   Topology & Routes & Runtime & Route No. & Runtime  & $\cU$ & Route No.  & Runtime & $\cU$ \\
   \hline
     \cref{fig:contention}(a) & 1024 & 0.255~s & 20 & 0.241~s & $-1.35$ & 20 & 0.255~s & $-1.35$ \\
   \cref{fig:combined_results_ring}(a)  & 10,000 & 136~s & 1 & 0.498~s & $73.2$  & 1203 & 62.8~s & $73.7$ \\
  \cref{fig:combined_results_dense}(a)  & 10,000 & 618~s & 1 & 0.283~s%0.599~s 
  & $51.1$  & 245 & 15.6~s & $51.2$
  \\ \hline 
    \end{tabular}
    \caption{Comparing pipeline performance to a semi-exhaustive search over up to 10,000 $(N^2S)^L$ possible lightpath combinations. The first column lists the network topology from \crefrange{fig:contention}{fig:combined_results_dense}, while the second column shows the total number of routes and runtime for a semi-exhaustive search. The route number, runtime, and utility of the first solution appear in the middle columns, while these values for the best solution found during the semi-exhaustive search appear in the rightmost columns.}
    \label{tab:optima}
\end{table}

\section{Discussion}
\label{sec:discussion}

Although global optimality cannot be \textit{guaranteed} without exhaustively checking all $(N^2S)^L$ possible double Yen combinations for sufficiently large $N$ and joint optimization of flux, channels, and bandwidth contention, our results' closeness to asymptotic maxima provides strong empirical evidence for feasibility.  %\rd{In about half a second, the ring network solution reached above $89\%$ of the infinite-resource maximum on average, and the Manhattan network reached above $93\%$ in about a quarter of a second.}  
A natural extension of our current methods would be to use more general link-dependent state and noise models instead of \cref{equ:fidelity,equ:rate}, which are tailored to all-optical networks based on spontaneous parametric downconversion \cite{alshowkan_broadband_2022} in which the dominant sources of noise stem from multipair emission \cite{takesue_effects_2010} and background counts, and where photons are modeled with a Werner-state approximation \cite{werner_quantum_1989, barbieri_generation_2004}. These extensions could incorporate more realistic channel effects such as polarization-mode dispersion \cite{rodimin_impact_2025} which has been shown to degrade broadband entanglement transmission, or even more environmentally based effects like time-varying background noise and polarization drift. More significantly, the direct entanglement viewpoint could be replaced by repeater models for end-to-end entanglement fidelity $\cF_\ell$ and rate $\cR_\ell$~\cite{vardoyan_quantum_2022, vardoyan_quantum_2023}. With these extensions, our three-stage pipeline would remain largely unchanged; all we would need to modify are \cref{equ:fidelity,equ:rate}.

An interesting consequence of large many-source, many-user networks is the potential for the lowest-loss paths to be suboptimal for global network utility. Intuitively, this situation occurs when the edge losses to some entanglement sources are much lower than to others, causing the double Yen algorithm in Phase~1 to overuse particular sources such that they are unable to allocate sufficient bandwidth to the collection of links assigned to them. At the same time, other sources positioned near high-loss edges may be underused or completely unused, suggesting missed opportunities to leverage their resources for higher global utility. Recognizing and responding to such false local optima will be critical for maximizing utility in large-scale dynamic quantum networks. In the near term, we plan to move in this direction by implementing the pipeline on physical flex-grid quantum networks \cite{alshowkan_resilient_2025, stevens_situ_2026} augmented with multiple sources to explore the ability to solve real-world RSA problems---particularly in cases where the optimal allocations may deviate from naive expectations.

\begin{backmatter}
\bmsection{Funding}
US Department of Energy (ERKJ432).

\bmsection{Acknowledgment}
This work was performed in part at Oak Ridge National Laboratory, operated by UT-Battelle for the US Department of Energy under contract DE-AC05-00OR22725. 

\bmsection{Disclosures}
The authors declare no conflicts of interest.

\bmsection{Data Availability}
Data and code used for all results in this paper are available at \url{https://github.com/zacharygoisman/QuantumNetwork}.

\end{backmatter}

\appendix
\section{Comparison to previous work}
\label{subsec:comparison}
In this appendix, we compare the performance of the pipeline proposed in the main text to the GA workflow of~\cite{alnas_optimal_2022}, applicable to star networks with a single entanglement source.
That paper provisioned resources to maximize a more complicated utility function (called ``fitness'' there):
\begin{equation}
\label{equ:previous_utility}
      \widetilde{\cU} = \sum_{\ell\in[L]}\left\{\frac{\cR_\ell\log_2 2\cF_\ell}{\max[\cR_\ell\log_2 2\cF_\ell]}\mathbbm{1}_{[f_\ell,1]}(\cF_\ell) - \mathbbm{1}_{[0,f_\ell)}(\cF_\ell) \right\},
\end{equation}
where $\cF_\ell$ and $\cR_\ell$ are defined in \cref{equ:fidelity,equ:rate}, $\mathbbm{1}_{[a,b]}(x)$ denotes the indicator function, and $\max[\cR_\ell\log_2 2\cF_\ell]$ is the ``infinite-resource'' value of $\cR_\ell\log_2 2\cF_\ell$---i.e., the maximum possible value of $\cR_\ell\log_2 2\cF_\ell$ in the limit of $K\rightarrow\infty$ frequency bin pairs. Since $\cR_\ell\log_2 2\cF_\ell$ equals the entangled bit rate, this utility function weights every link exceeding its fidelity threshold $f_\ell$ according to its fraction of the infinite-resource entanglement rate, while every link unable to reach its fidelity threshold penalizes the utility by $-1$. Significantly, because the penalty for $\cF_\ell< f_\ell$ is finite, it is possible that the utility-maximizing solutions might drop some links altogether, in contrast to the current utility function [\cref{equ:sumrate}] that penalizes $\cF_\ell<f_\ell$ with $\log_{10}0=-\infty$.

\begin{figure}[tb!]
\centering
\includegraphics[width=1\linewidth]{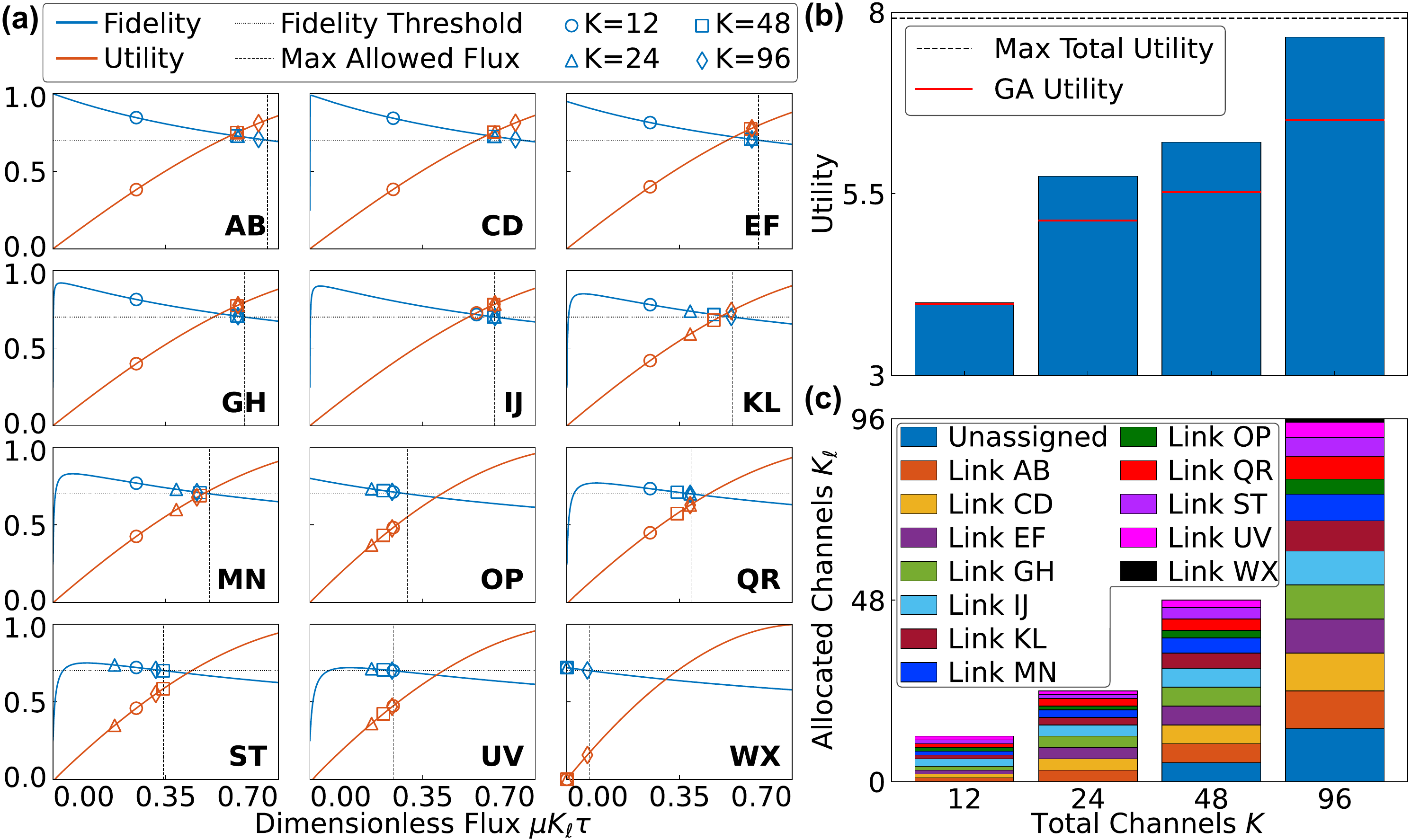}
\caption{APOPT optimization of Scenario 4 from \cite{alnas_optimal_2022}. Each combination of two letters represents an entangled link pair between two users. This $(S,U,L)=(1,24,12)$ star network has a fidelity constraint $f_\ell=0.7$ for all links. (a) Fidelity and utility [\cref{equ:fidelity,equ:previous_utility}] with the best solution at each $K$ denoted by the relevant symbol. (b) Total utility of APOPT (blue bars) compared to the GA utility from \cite{alnas_optimal_2022} (red lines) and infinite-resource limit $K\rightarrow\infty$ (black dashed line). (c) Channel number allocations $K_\ell$ for each APOPT solution.
\label{fig:compare4}}
\end{figure}

In what follows, we consider this previous utility function $\widetilde{\cU}$ to compare our APOPT spectrum allocation method's effectiveness to \cite{alnas_optimal_2022}. Results for the most complex case there---Scenario 4 with $L=12$ links and $f_\ell\geq0.7$ (Fig.~7 in \cite{alnas_optimal_2022})---are shown in \cref{fig:compare4}, with (a) highlighting the APOPT-achieved fidelity and utility for each link as a function of allocated flux under  available  channel pairs $K$, (b) presenting the total utility for all links, and (c) showing the number of channels each link is allocated. 

Significantly, APOPT's utility values are comparable to or even higher than GA's, with \cref{fig:compare4}(b) showing improved utilities $\widetilde{\cU}_\text{APOPT}-\widetilde{\cU}_\text{GA}\in\{0.61,0.69,1.17\}$ for $K\in\{24, 48, 96\}$.
Yet even more remarkable is the speed with which APOPT finds these solutions---namely, $\{0.36~\text{s},0.61~\text{s},1.44~\text{s},1.07~\text{s}\}$ for $K\in\{12,24,48,96\}$, compared to $\{24~\text{s},59~\text{s},43~\text{s},68~\text{s}\}$ for GA on the same computer. Therefore, the APOPT solver demonstrates that it can both reach and even improve upon the spectrum allocation done by GA in significantly less time.

\bibliography{bibliography.bib}

\end{document}